\documentclass{CHEP2006}
\usepackage{graphicx}
\usepackage{hyperref}
\usepackage{url}
\begin{document}
\title{CMS Software Distribution on the LCG and OSG Grids}

\author{
  K.~Rabbertz, University of Karlsruhe, Karlsruhe, Germany\\
  M.~Thomas, CALTECH, Pasadena, California, USA\\
  S.~Ashby, CERN, Geneva, Switzerland\\
  M.~Corvo, CERN \& INFN, Padova, Italy\\
  S.~Argir\`o, CERN \& INFN-CNAF, Bologna, Italy\\
  N.~Darmenov, CERN \& INRNE, Sofia, Bulgaria\\
  R.~Darwish, D.~Evans, B.~Holzman, N.~Ratnikova, FERMILAB, Batavia, Illinois , USA\\
  S.~Muzaffar, Northeastern University, Boston, Massachusetts, USA\\
  A.~Nowack, RWTH Aachen University, Aachen, Germany\\
  T.~Wildish, Princeton University, Princeton, New Jersey , USA\\
  B.~Kim, University of Florida, Gainesville, Florida, USA\\
  J.~Weng, University of Karlsruhe \& CERN, Karlsruhe, Germany\\
  V.~B\"uge, University of Karlsruhe \& FZK, Karlsruhe, Germany}

\maketitle

\begin{abstract}
  The efficient exploitation of worldwide distributed storage and
  computing resources available in the grids require a robust,
  transparent and fast deployment of experiment specific software. The
  approach followed by the CMS experiment at CERN in order to enable
  Monte-Carlo simulations, data analysis and software development in
  an international collaboration is presented. The current status and
  future improvement plans are described.
\end{abstract}

\section{INTRODUCTION\label{SEC:CHEP247-INTRO}}

The CMS (Compact Muon Solenoid) experiment~\cite{CMS-PTDRI,CMS-web} of
high energy physics is located in an underground cavern at the Large
Hadron Collider (LHC) currently under construction at CERN, Geneva,
Switzerland~\cite{LHC-web}. The international collaboration tackling
the difficult task of constructing and running the complex detector
weighing 12.5 tons comprises about 2000 scientists and engineers from
160 institutions in 37 countries.  With an expected recording rate of
150 events per second and event sizes of about $1.5\,{\rm MB}$ the
huge amount of $1500\,{\rm TB}$ of collected data per year has to be
distributed on grids in order to be stored and analyzed.  The
efficient exploitation of the worldwide distributed data and the
computing resources available in the grids require a robust,
transparent and fast deployment of the experiment specific software
that, especially in the start-up phase, will be rapidly developing.
The approach followed by the CMS experiment in order to achieve this
goal is presented, the current status of the implementations within
the LHC Computing Grid (LCG)~\cite{LCG-TDR,LCG-web} and the Open
Science Grid (OSG)~\cite{OSG:CHEP06-423,OSG-web} and future
improvement plans are described.

\section{SOFTWARE PREPARATION\label{SEC:CHEP247-SOFTPREP}}

Before the rapidly developing experiment software can be distributed a
couple of preparatory steps have to be performed. They are listed in
the following together with a short explanation and, if applicable,
the solution adopted by CMS:
\begin{enumerate}
\item {\bf Release:} The final content of a project has to be fixed.
  Within CMS the software projects are managed by SCRAM (Software
  Configuration, Release And Management)~\cite{Wellisch:2003wb}, the
  collection of the latest updates is done via NICOS (Nightly Control
  System)~\cite{Undrus:2003xr}. See also
  reference~\cite{Argiro:CHEP06-246}.
\item {\bf Packaging:} All software components have to be packaged in
  archives suited for distribution on grids. CMS adopted the RPM (Red
  Hat Package Manager)~\cite{MaximumRPM} format, see
  also~\cite{Nowack:CHEP06-248}.
\item {\bf Testing:} A test installation has to be performed followed
  up by a validation.
\item {\bf Archiving:} All produced packages have to be backed up, for
  CMS they are stored on tapes managed by CASTOR (CERN Advanced
  Storage Manager)~\cite{Baud:2003ys,CASTOR-web}.
\item {\bf Web/Grid Storage:} The validated software archives have to
  be put into a repository accessible by web or grid tools.  CMS
  employs a web server for this task, in addition the archived copies
  on CASTOR can be accessed via grid tools.
\item {\bf Publication:} New releases ready for distribution have to
  be announced. This is done using the same web server as for the
  repository.
\item {\bf Mirroring:} Ideally, to avoid overloading the primary
  repository, mirrors should be set up.
\end{enumerate}

\section{SOFTWARE DISTRIBUTION\label{SEC:CHEP247-SOFTDIST}}

\subsection{Generic View\label{SEC:CHEP247-GENVIEW}}

\begin{figure*}[t]
  \centering \includegraphics*[width=145mm]{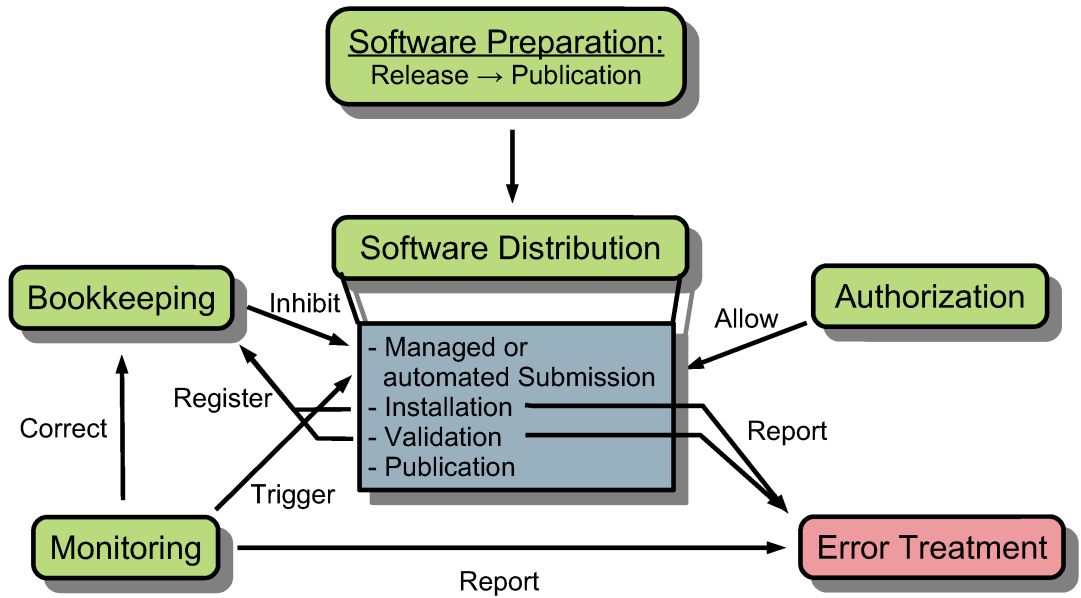}
  \caption{A generic view showing the required services in order to
    prepare, distribute and maintain experiment specific software on a
    grid. Labelled arrows indicate their interconnections, the
    rectangular box provides some more details on the software
    distribution service.}
  \label{fig:chep247-GenericView}
\end{figure*}

Once the software has been prepared and the required services have
been set up the actual distribution on the grids can start. In
figure~\ref{fig:chep247-GenericView} a generic view on the related
services and their interconnections is presented. The software
distribution service, where the rectangular box provides some more
details, consists of the four basic steps: Submission, installation,
validation and publication. In comparison to local software
installations additional grid services come into play.  To avoid
misuse or unintended destructive actions the submission, which might
be initiated in a managed or automated manner, has to be authorized
for access to the software storage area. In addition, information
retrieved from a bookkeeping service can prevent unwanted (or double)
submissions.  The next two steps are similar to the point "Testing" of
the software preparation section with the added complication that the
actions have to be registered. In parallel to the distribution of new
releases the current status of all participating grid sites is
monitored so that changes in availability or e.g.\ deteriorated
systems can be reported to an error treatment service and the entries
in the bookkeeping are updated correspondingly. At the same time the
monitoring can be used to trigger installation submissions of new
releases in an automated manner. Any other problems occurring for
example in the installation or validation phases are reported to the
error treatment as well.

\subsection{Implementation within LCG\label{SEC:CHEP247-LCGIMP}}

Within the LHC Computing Grid the whole chain of submission,
installation, validation and publication is performed by the tool
XCMSI~\cite{Rabbertz:CHEP04-137,XCMSI-web}, the publication occurs
according to the GLUE (Grid Laboratory Uniform Environment)
scheme~\cite{LCG-TDR,GLUE-web}. XCMSI also comprises a monitoring
service allowing automated submissions but no dedicated bookkeeping
apart from a simple web page, more details can be found in the section
on monitoring and bookkeeping. The submissions are authorized using
X509 grid certificates of the mapping account {\tt cmssgm} of the
experiment software manager (ESM). The only means of error treatment
currently implemented is the Savannah web page of the XCMSI
project~\cite{XCMSI-Savannah}.

\subsection{Implementation within OSG\label{SEC:CHEP247-OSGIMP}}

In the Open Science Grid the software installations are submitted from
the CMS Software Deployment GUI (Graphical User Interface) by an
experiment software manager again authorized by X509 grid certificates
for the role of {\tt cmssoft}. The installation is done with XCMSI
like within LCG, to validate the software, however, a series of
Monte-Carlo production jobs is run employing the Monte-Carlo
Production Service (MCPS)~\cite{MCPS-web}. The result is published in
the GLUE scheme as well as in the dedicated bookkeeping database
CMSSoftDB~\cite{CMSSoftDB-web}.  A continuous monitoring has not been
foreseen, the only available error treatment is again the XCMSI
Savannah page.

\subsection{Comparison\label{SEC:CHEP247-COMP}}

In general the experience from the Data Challenge 04~\cite{CMS-CTDR}
has lead to rather similar setups of the software deployment on the
LCG and OSG grids. Some components, especially the validation
procedures and the escalation handling in case of problems, have to be
further developed.

\section{MONITORING AND BOOKKEEPING\label{SEC:CHEP247-MONBOOK}}

As can be seen from figure~\ref{fig:chep247-GenericView} the task of
software deployment is not considered to be completed after just one
initial distribution of a new release. The current status has to be
monitored continuously in order to update the published information on
grid sites according to the actual situation and to take preventive
measures in case of problems. Within LCG the basic availability of a
grid site and the correct functioning of the corresponding compute and
storage elements is already monitored in the framework of the site
functional tests (SFT). In principal it is possible to add experiment
specific tests on the installed software as well which would have the
additional advantage that the jobs are run in privileged (express)
queues. If a validation, however, requires that files within the
experiment software area are modified or created it would be necessary
that not only the experiment software manager but also the SFTs run by
the LCG integration team (dteam) are authorized to have write access.
The same would be required if the monitoring is not only run in a
passive testing mode but also in active mode that allows known
problems to be fixed automatically or that even submits installation
jobs for new releases.  Since this is not very desirable a better
solution should be found to prioritize software monitoring and
installation jobs of the experiment software manager with respect to
normal ones. This could be implemented for example in the framework of
the Virtual Organization Membership Service (VOMS) where it is
possible to attribute different authorizations and prioritizations to
a grid certificate depending on the role the submitter
assumes~\cite{LCG-TDR}.

The aforementioned XCMSI project contains a monitoring tool that is
already run regularly (in unprivileged queues though) in order to run
CMS specific tests e.g.\ to check the read/write permissions of the
experiment software area, the CMS attributed architecture (operating
system) of the compute element, the availability of the RPM database
of installed packages and the accessibility of the published software
projects. The collected data are published in the form of a web page
presenting also some additional information on the outcome of the last
test and a history file containing the past test results.

A dedicated database to gather all the data in a nicely structured
format is desirable. This would also improve the capabilities of the
automated installation prototype contained in the XCMSI monitoring
that currently relies on GLUE tag information like {\tt
  VO-cms-PROJECT-request-install} alone. Such a database has been
implemented within the OSG: CMSSoftDB~\cite{CMSSoftDB-web}. It is
based on a MySQL~\cite{MySQL-web} database and provides i.a\@. a
comprehensive overview of the CMS software installation status in the
Open Science Grid. It is accessible via a web interface, the CMS
Software Deployment GUI~\cite{Kim:APS-CB.00003,CSDGUI-web}, that not
only presents the collected information in the database but also
allows authorized users to perform actions like job submissions or
other management tasks. The database is not coupled to a continuous
monitoring though.

\section{OUTLOOK\label{SEC:CHEP247-OUTLOOK}}

A lot of progress has been made compared to the situation during the
Data Challenge 04. CMS can deploy the experiment specific software,
distribute and analyze data and monitor the software status on the
grids. The interoperability between LCG and OSG has been improved as
well but some work remains to be done, especially a more consistent
look onto the available information for users in the two grids should
be achieved. Concerning monitoring and bookkeeping the efforts made
within LCG and OSG could be nicely merged. Dedicated CPU/time slots
for the ESM role are a must, but this can easily be done within the
VOMS roles. The most important points remaining to be addressed are the
error handling and escalation procedures as well as better validation
suites.

%
% BibTeX Bibliography
%
\bibliographystyle{lucas_unsrt} \bibliography{chep06-rabbertz-247}

\end{document}